\documentclass{article}
\usepackage{spconf,amsmath,graphicx}
\usepackage{amssymb}
\usepackage{color}

\usepackage{caption}
\usepackage{subcaption}

\def\0{{\mathbf 0}}
\def\1{{\mathbf 1}}

\def\e{{\mathbf e}}

\def\u{{\mathbf u}}
\def\v{{\mathbf v}}

\def\x{{\mathbf x}}
\def\y{{\mathbf y}}

\def\A{{\mathbf A}}
\def\B{{\mathbf B}}
\def\C{{\mathbf C}}
\def\D{{\mathbf D}}
\def\E{{\mathbf E}}

\def\L{{\mathbf L}}

\def\T{{\mathbf T}}
\def\U{{\mathbf U}}

\def\W{{\mathbf W}}

\def\Y{{\mathbf Y}}

\def\ie{{\textit{i.e.}}}

\def\cE{{\mathcal E}}

\def\cG{{\mathcal G}}
\def\cH{{\mathcal H}}

\def\cL{{\mathcal L}}

\def\cS{{\mathcal S}}

\def\cV{{\mathcal V}}

\def\balpha{{\boldsymbol \alpha}}

\def\bzeta{{\boldsymbol \zeta}}

\title{Hybrid Model-based / Data-driven Graph Transform for Image Coding}
\name{
Saghar Bagheri$^{\dag}$, 
Tam Thuc Do$^\dag$\thanks{Equal contributions from the first and second authors.}, 
Gene Cheung$^\dag$\thanks{Gene Cheung acknowledges the support of the NSERC grants RGPIN-2019-06271,  RGPAS-2019-00110.}, 
Antonio Ortega$^\star$}
\address{
  $^\dag$York University, Toronto, Canada\\ 
  $^\star$University of Southern California, CA, USA}
\begin{document}
\ninept
\maketitle
\begin{abstract}
Transform coding to sparsify signal representations remains crucial in an image compression pipeline.
While the Karhunen-Lo\`{e}ve transform (KLT) computed from an empirical covariance matrix $\bar{\C}$ is theoretically optimal for a stationary process, in practice, collecting sufficient statistics from a non-stationary image to reliably estimate $\bar{\C}$ can be difficult. 
In this paper, to encode an intra-prediction residual block, we pursue a hybrid model-based / data-driven approach: the first $K$ eigenvectors of a transform matrix are derived from a statistical model, e.g., the asymmetric discrete sine transform (ADST), for stability, 
while the remaining $N-K$ are computed from $\bar{\C}$ for performance.
The transform computation is posed as a graph learning problem, where we seek a graph Laplacian matrix minimizing a graphical lasso objective inside a convex cone sharing the first $K$ eigenvectors in a Hilbert space of real symmetric matrices. 
We efficiently solve the problem via augmented Lagrangian relaxation and proximal gradient (PG).
Using WebP as a baseline image codec, experimental results show that our hybrid graph transform achieved better energy compaction than default discrete cosine transform (DCT) and better stability than KLT. 
\end{abstract}
\section{Introduction}
\label{sec:intro}
Transform coding remains a fundamental component in a conventional image / video compression pipeline: an input pixel block $\x \in \mathbb{R}^N$ is transformed to a sparse representation $\balpha = \T \x$, where $\T \in \mathbb{R}^{N \times N}$ is the transformation matrix, before scalar quantization and entropy coding, resulting in good coding performance \cite{goyal01}.
While \textit{fixed} transforms such as \textit{discrete cosine transform} (DCT) \cite{strang99} derived from statistical models are used traditionally, \textit{adaptive} transforms that are computed on-the-fly based on an image's local statistics often perform better thanks to its adaptability.  
Specifically, \textit{Karhunen-Lo\`{e}ve transform} (KLT) \cite{pearlman08} computed from an empirical covariance matrix $\bar{\C} \in \mathbb{R}^{N \times N}$ is theoretically optimal, in terms of signal decorrelation, for a stationary process.
However, in practice, collecting sufficient statistics from images that are known to be non-stationary to reliably estimate $\bar{\C}$ can be difficult.
This means that the computed KLT---composed of eigenvectors of $\bar{\C}$---can become unstable, leading to an unacceptably large variance of coding performance across different blocks. 

To alleviate this problem, leveraging on our previous work on spectral graph learning\footnote{While \cite{bagheri21} focuses on the computation of graph Laplacian $\L$ with pre-chosen first $K$ eigenvectors given empirical covariance $\bar{\C}$, we study the learning and deployment of $\L$ for hybrid transform coding of images.} \cite{bagheri21}, we pursue a \textit{hybrid model-based / data-driven approach} to code an intra-prediction residual block, where the first $K$ eigenvectors of a transform matrix are determined by a statistical model such as \textit{asymmetric discrete sine transform} (ADST) \cite{han12} for stability, while the remaining $N-K$ eigenvectors are computed from $\bar{\C}$ for performance.
Unique in this design is that the parameter $K$ is \textit{tunable} and can be chosen depending on the reliability of estimated $\bar{\C}$: 
if $\bar{\C}$ is deemed unreliable due to insufficient training data, then a larger $K$ is chosen, so that fewer (and less important) eigenvectors are computed from $\bar{\C}$.

Mathematically, we compute the hybrid transform as a constrained graph learning problem: 
we first define a convex cone $\cH^+_{\u}$ in a Hilbert space \cite{vetterli2014foundations} of symmetric real matrices that share the first $K$ eigenvectors $\{\u_k\}_{k=1}^K$, then seek an optimal graph Laplacian matrix $\L \in \cH^+_{\u}$ that minimizes a \textit{graphical lasso} (GLASSO) objective \cite{friedman08} given $\bar{\C}$.  
We show that optimizing each one of $N-K$ remaining eigenvectors from $\bar{\C}$ is NP-hard, but we efficiently approximate the problem using augmented Lagrangian relaxation \cite{Lagrangian} and \textit{proximal gradient} (PG) \cite{parikh13}. 
Using Google WebP as a baseline image scodec and focusing on coding prediction residuals from intra-prediction mode DC4\footnote{Statistics show that DC4 mode is used approximately $80\%$ of the time among available intra-prediction modes.}, experimental results using standard test images show that our hybrid graph transform achieved better energy compaction than default DCT and better stability than KLT in terms of variation from average performance.

The outline of the paper is as follows. 
We first review GSP definitions and GLASSO in Section\;\ref{sec:prelim}. 
We outline our transform impelmentation in WebP in Section\;\ref{sec:webp}. 
We describe our hybrid transform optimization in Section\;\ref{sec:learn}.
Experimental results and conclusion are presented in Section\;\ref{sec:results} and \ref{sec:conclude}, respectively.

\vspace{0.05in}
\noindent 
\textbf{Related Works}:
The impracticality of KLT in speed and memory requirements was addressed in numerous works \cite{pirooz98,greeshields98,blanes11}. 
We focus instead on the problem of insufficient statistics when coding non-stationary images, and propose a hybrid model-based / data-driven transform towards a good tradeoff in compression performance and stability.
Compression using graph transforms \cite{hu15,hu15spl,su17,egilmez20,chao22}---transformation of signals on graphs from the nodal domain to the graph spectral domain---have been studied in the \textit{graph signal processing} (GSP) literature \cite{ortega18ieee,cheung18} during the past decade.
Our graph transform is unique in that our GLASSO-based optimization allows the optimal combination of $K$ model-based eigenvectors with $N-K$ remaining ones computed from data, resulting in a stable transform.

\section{Preliminaries}
\label{sec:prelim}
\subsection{GSP Basics}
\label{subsec:GSP}

A graph $\cG(\cV,\cE,\W)$ is defined by a set of $N$ nodes $\cV = \{1, \ldots, N\}$, edges $\cE = \{(i,j)\}$, and an \textit{adjacency matrix} $\W$. 
$W_{i,j} \in \mathbb{R}$ is the edge weight if $(i,j) \in \cE$, and $W_{i,j} = 0$ otherwise. 
Self-loops may exist, in which case $W_{i,i} \in \mathbb{R}^+$ is the weight of the self-loop for node $i$.
\textit{Degree matrix} $\D$ has diagonal entries $D_{i,i} = \sum_{j} W_{i,j}, \forall i$. 
A \textit{combinatorial graph Laplacian matrix} $\L$ is defined as $\L \triangleq \D - \W$, which is provably \textit{positive semi-definite} (PSD) for positive graphs \cite{cheung18}. 
If self-loops exist, then the \textit{generalized graph Laplacian matrix} $\cL$, defined as $\cL \triangleq \D - \W + \text{diag}(\W)$, is often used.
Any real symmetric matrix can be interpreted as a generalized graph Laplacian matrix.

\subsection{Hilbert Space Definitions}

We first define a \textit{vector space} $\cS$ of real, symmetric matrices in $\mathbb{R}^{N \times N}$. 
We next define an inner product $\langle \cdot, \cdot \rangle$ for two matrices $\A, \B \in \cS$ as
\begin{align}
\langle \A, \B \rangle = \text{tr}(\B^{\top} \A) = \sum_{i,j} A_{ij} B_{ij}.
\end{align}
Assuming Cauchy sequence convergence, the vector space endowed with an inner product is a \textit{Hilbert Space} $\cH$ \cite{vetterli2014foundations}. 
We define a \textit{subspace} $\cH^+ \subset \cH$ that contains PSD matrices, \ie, $\cH^+ = \{\A \in \cH \,|\, \A \succeq 0\}$.
It can be easily proven that $\cH^+$ is a convex cone \cite{bagheri21thesis}.
Further, define $\cH^+_{\u} \subset \cH^+$ as the subset of matrices that share the first $K$ eigenvectors $\{\u_k\}_{k=1}^K$. 
$\cH_{\u}^+$ can also be proven to be a convex cone \cite{bagheri21thesis}; this implies that projection to $\cH_{\u}^+$ is a projection to a convex set.

\subsection{Graphical Lasso}
\label{subsec:glasso}

Given an empirical covariance matrix $\bar{\C}$ estimated from data, GLASSO formulates the following problem for inverse covariance (precision) matrix $\L$:
\begin{align}
\min_{\L \in \cH^+} \text{Tr}(\L \bar{\C}) - \log \text{det} \L + \rho \|\L\|_1 .
\label{eq:GLASSO}
\end{align}
The first two terms in \eqref{eq:GLASSO} can together be interpreted as the likelihood given observation $\bar{\C}$, while the last $\ell_1$-norm term promotes sparse reconstruction of $\L$. 
As done in \cite{egilmez17}, we interpret computed $\L$ as a generalized graph Laplacian matrix, and thus \eqref{eq:GLASSO} can be considered a graph learning formulation given input covariance $\bar{\C}$.  

\section{WebP}
\label{sec:webp}
\begin{figure}[t]
\centering
\begin{subfigure}[b]{0.2\textwidth}
\centering
\includegraphics[width=\textwidth]{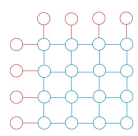}
\caption{\small Intra-prediction}
\label{fig:intra_prediction}
\end{subfigure}
\hfill
\begin{subfigure}[b]{0.25\textwidth}
\centering
\includegraphics[width=\textwidth]{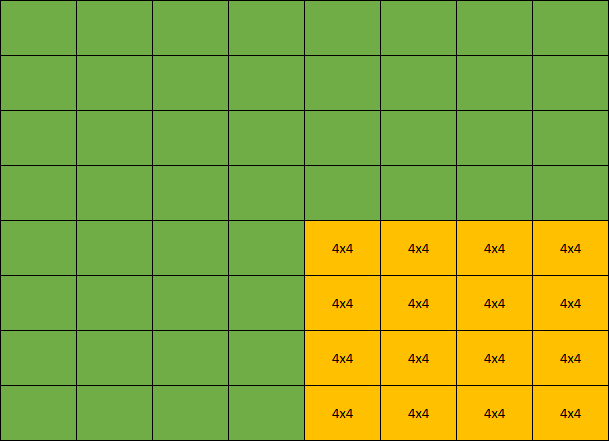}
\caption{\small Blocks and sub-blocks}
\label{fig:CovarData}
\end{subfigure}
\vspace{-0.05in}
\caption{\small DC4 intra-prediction mode in WebP: (a) a target $4 \times 4$ pixel sub-block (blue) is intra-predicted using $8$ reference pixels (red); (b) $4 \times 4$ sub-blocks inside a $16 \times 16$ target block (yellow) are intra-predicted and coded individually using DCT. In our implementation, neighboring $16 \times 16$ decoded blocks (green) are used to estimate covariance matrix $\bar{\C}$ for our hybrid graph transform computation. }
\label{fig:dc4mode}
\end{figure}

We use Google's open source WebP\footnote{https://developers.google.com/speed/webp} as a baseline image codec on which we implement our hybrid graph transform for transform coding of intra-prediction residuals.
Specifically, we focus on prediction residuals of the DC4 intra-prediction mode, illustrated in Fig.\;\ref{fig:dc4mode}(a). 
To predict a target $4 \times 4$ sub-block of pixels (blue), the eight adjacent horizontal and vertical coded pixels (red) are used as reference to compute an average value $p$.
We subtract $p$ from the ground truth target sub-block for a $4 \times 4$ prediction residual, which is transform-coded using DCT in WebP.

In our hybrid transform implementation, instead of DCT, to encode each $4 \times 4$ sub-block inside a $16 \times 16$ target block (yellow in Fig.\;\ref{fig:dc4mode}(b)), we first compute an empirical covariance matrix $\bar{\C}$ using three neighboring $16 \times 16$ coded blocks (green) as follows. 
Using the reference coded pixels, we first mimic DC4 intra-prediction and compute $M$ prediction residuals $\{\y_m\}_{m=1}^M$, where $\y_m \in \mathbb{R}^{16}$ is the $m$-th residual. 
Covariance is then computed empirically as $\bar{\C} = \frac{1}{M} \sum_{m=1}^M \y_m \y_m^\top$. 
The same $\bar{\C}$ is used for transform coding of all $4 \times 4$ sub-blocks inside the target $16 \times 16$ block.

\section{Learning Graph Transform}
\label{sec:learn}
Given an empirical covariance matrix $\bar{\C}$ as previously discussed, we now compute a hybrid graph transform.  
We first develop a \textit{projection operator} $\text{Proj}(\cdot)$ to project $\bar{\C}^{-1}$ to convex cone $\cH_{\u}^+$, where $\{\u_k\}_{k=1}^K$ are the first $K$ orthonormal eigenvectors derived \textit{a priori} from a statistical model like ADST \cite{han12}.
We seek the optimal transform via a modified GLASSO formulation, where we use our developed $\text{Proj}(\cdot)$ in an iterative algorithm to compute a solution.

\subsection{Eigen-Pair $(\lambda_1,\u_1)$}

Given covariance $\bar{\C}$, the operator $\text{Proj}(\cdot)$ computes one eigen-pair at a time to compose Laplacian $\L = \sum_{k=1}^K \lambda_k \u_k \u_k^\top + \sum_{i=K+1}^N \lambda_i \v_i \v_i^\top$, 
where $\{\u_k\}_{k=1}^K$ and $\{\v_i\}_{i=K+1}^N$ are the model-based and data-driven orthonormal eigenvectors, respectively. 
Operator $\text{Proj}(\cdot)$ is a projection \cite{vetterli2014foundations} since it is provably \textit{idempotent}, \ie, $\text{Proj}(\text{Proj}(\A)) = \text{Proj}(\A), \forall \A \in \cH$ \cite{bagheri21thesis}.

We first compute the \textit{first} eigen-pair $(\lambda_1,\u_1)$ for $\L$, or equivalently, the \textit{last} eigen-pair $(\mu_N = 1/\lambda_1, \u_1)$ for $\C = \L^{-1}$, where $\u_1$ is a known model-based eigenvector. 
Specifically, we project $\bar{\C}$ onto 1D subspace spanned by rank-1 matrix $\u_1 \u_1^{\top}$ to maximally preserve $\bar{\C}$.
This results in $\mu_N$:
\begin{align}
\mu_N = \langle \bar{\C}, \u_1 \u_1^\top \rangle .
\label{eq:first_u}
\end{align}
We compute residual signal $\E_{N}$ as
\begin{align}
\E_{N} = \bar{\C} - \langle \bar{\C}, \u_1 \u_1^\top \rangle \u_1 \u_1^\top .
\label{eq:first_E}
\end{align}

\subsection{Eigen-Pair $(\lambda_{k},\u_k), k \in \{2, \ldots, K\}$} 

For eigen-pair $(\lambda_k,\u_k)$ of $\L$, $k \in \{2, \ldots, K\}$, where $\u_k$ is also known, we compute the $(N-k+1)$-th eigenvalue $\mu_{N-k+1} = 1/\lambda_k$ of $\C$ as
\begin{align}
\mu_{N-k+1} = \min \left( \langle \E_{N-k+2}, \u_k \u_k^\top \rangle, \mu_{N-k+2} \right). 
\label{eq:next_u}
\end{align}
\eqref{eq:next_u} is similar to \eqref{eq:first_u}, with the addition of the minimization to ensure the ordered eigenvalues $\{\mu_i\}_{i=1}^N$ are non-decreasing.

\subsection{Eigen-Pair $(\lambda_{i}, \v_i), i \in \{K+1, \ldots, N\}$}

For eigen-pair $(\lambda_{i}, \v_i)$ of $\L$, $i \in \{K+1, \ldots N\}$, where unknown $\v_i$ needs to be computed from data, we seek a unit-norm $\v_i$ that maximizes the following inner product to maximally preserve residual signal $\E_{N-k+2}$:
\begin{align}
\v_{i} &= \arg \max_{\v} ~~ \langle \E_{N-k+2}, \v \v^{\top} \rangle, 
\nonumber \\
&\mbox{s.t.}~ \left\{
\begin{array}{l}
\u_k^\top \v = 0, ~~\forall k \in \{1, \ldots, K\} \\
\v_j^\top \v = 0, ~~\forall j \in \{K+1, \ldots, i-1\} \\
\|\v\|_2 = 1
\end{array} 
\right. .
\label{eq:next_v}
\end{align}
The constraints require $\v_i$ to be orthogonal to the first $K$ known eigenvectors $\{\u_k\}_{k=1}^K$ and the previously computed $\{\v_j\}_{j=K+1}^{i-1}$.
Objective \eqref{eq:next_v} is equivalent to $\v^{\top} \E_{N-k+2} \v$, which is quadratic and convex, given $\E_{N-k+2}$ can be proven to be PSD \cite{bagheri21thesis}.
Thus, maximization in \eqref{eq:next_v} is non-convex and NP-hard.



\subsubsection{Fast Approximation}

We perform a fast approximation for \eqref{eq:next_v}. 
We first approximate $\E_{N-k+2}$ with its rank-1 approximation $\e \e^{\top}$, where $\e$ is the last eigenvector\footnote{Extreme eigenvectors can be computed in roughly linear time using fast algorithms like LOBPCG \cite{lobpcg}.} of $\E_{N-k+2}$.
We then formulate the following problem:
\begin{align}
\max_{\v} \e^{\top} \v,    
~~\mbox{s.t.}~
\left\{ \begin{array}{l}
\u_k^{\top} \v = 0, ~~\forall k \in \{1, \ldots, K\} \\
\v_j^{\top} \v = 0, ~~\forall j \in \{K+1, \ldots, i-1\} \\
\| \v \|_2^2 \leq 1
\end{array} \right. .
\label{eq:approx1}
\end{align}
The first two constraint are the same as \eqref{eq:next_v}. 
The third constraint is a relaxation of $\|\v\|_2 = 1$ in \eqref{eq:next_v}, so that the feasible solution space is a convex set.

Then, we rewrite the constrained problem \eqref{eq:approx1} into the corresponding unconstrained version as follows.
We first define $\Y \in \mathbb{R}^{N \times (N-i)}$ as a matrix containing eigenvectors to-date:
\begin{align}
\Y = \left[\u_1 \cdots \u_K \v_{K+1} \cdots \v_{i-1} \right] .
\end{align}
We next define the \textit{augmented Lagrangian} \cite{Lagrangian} of the constrained problem \eqref{eq:next_v} as
\begin{align}
\min_{\v} - \e^{\top} \v + \bzeta^\top \Y^\top \v + \gamma \v^{\top} \Y \Y^{\top} \v + \Phi_1(\v)
\label{eq:approx2}
\end{align}
where $\bzeta \in \mathbb{R}^{N-i}$ is the Lagrange multiplier vector, and $\gamma > 0$ is a weight parameter.
$\Phi_1(\v)$ is a convex function defined as
\begin{align}
\Phi_1(\v) = \left\{ \begin{array}{ll}
0 & \mbox{if}~~ \|\v\|_2^2 \leq 1 \\
\infty & \mbox{o.w.} 
\end{array} \right. .
\end{align}
Denote by $\Theta(\v) = -\e^{\top} \v + \bzeta^\top \Y^\top \v + \gamma \v^{\top} \Y \Y^{\top} \v$.
Thus, the objective in \eqref{eq:approx2} is composed of two functions: 
i) $\Theta(\v)$ is convex and differentiable w.r.t. $\v$ with gradient $\nabla \Theta(\v) = - \e + \Y \bzeta + 2 \gamma \Y \Y^\top \v$, and
ii) $\Phi_1(\v)$ is convex and non-differentiable. 
One can thus optimize variable $\v$ in \eqref{eq:approx2} iteratively using \textit{proximal gradient} (PG) \cite{parikh13}, where multiplier $\zeta_k^{t+1}$ at iteration $t+1$ is updated using $\zeta_k^t$ and solution $\v^t$ at iteration $t$ as

\begin{small}
\begin{align}
\zeta_k^{t+1} = \left\{ \begin{array}{ll}
\zeta_k^{t} + \gamma \u_k^{\top} \v^t & \mbox{if} ~ k \in \{1, \ldots, K\} \\
\zeta_k^{t} + \gamma \v_k^{\top} \v^t & \mbox{if} ~ k \in \{K+1, \ldots, i-1\}
\end{array} \right. .
\end{align}
\end{small}
For initialization, we set the first solution to be $\v^0 = \e/\|\e\|^2_2$.

\subsubsection{Compute Eigenvalue $\lambda_{i}$}

Given approximated eigenvector $\v_{i}$ from data, we compute the corresponding eigenvalue $\mu_{N-i+1} = 1/\lambda_{i}$ as
\begin{align}
\mu_{N-k+1} = \min \left( \langle \E_{N-k+2}, \v_{i} \v_{i}^{\top} \rangle, \mu_{N-k+2} \right)  .
\label{eq:next_eigenvalue}
\end{align}
Residual is updated as $\E_{N-i+1} = \E_{N-i+2} - \mu_{N-k+1} \v_{i} \v_{i}^{\top}$.

\subsection{Modified GLASSO Formulation}

The previous eigen-component computation constitutes a projection $\L = \text{Proj}(\bar{\C}^{-1})$.  
We now formulate the following GLASSO-like optimization problem to estimate a graph Laplacian matrix $\L$ \cite{mazumder2012graphical}:
\begin{align}
\min_{\L \in \cH^+_{\u}} ~~ \text{Tr}(\L \bar{\C}) -\log \det \L + \rho \; \| \L \|_1
\label{eq:glasso}
\end{align}
where $\rho > 0$ is a shrinkage parameter for the $l_1$-norm.
The only difference from GLASSO is that \eqref{eq:glasso} has an additional constraint $\L \in \cH^+_{\u}$. 

We solve \eqref{eq:glasso} iteratively using projection operator $\text{Proj}(\cdot)$ and a variant of the \textit{block Coordinate descent} (BCD) algorithm in \cite{wright2015coordinate}. 
Specifically, we solve the \textit{dual} of GLASSO as follows.
Note first that the $l_1$-norm in \eqref{eq:glasso} can be written as 
\begin{align}
  \|\L\|_1 = \max_{\| \U \|_\infty\leq 1} ~~\text{Tr}(\L\U) 
\end{align}
where $\| \U \|_\infty$ is the maximum absolute value element of matrix $\U$. 
Hence, the dual problem of GLASSO that seeks an estimated covariance matrix $\C = \L^{-1}$ is
\begin{align}
\min_{\C \in \cH^+_{\u}} ~~ -\log \det \C, 
~~~\mbox{s.t.} ~~ 
\| \C - \bar{\C} \|_\infty \leq \rho
\label{eq:glasso_dual}    
\end{align}
where $\C =\bar{\C}+\U$  implies that the primal and dual variables are related via $\L = {(\bar{\C}+\U)}^{-1} $ \cite{Banerjee7}.
The unconstrained objective in \eqref{eq:glasso_dual} can be iteratively minimized by updating one row-column pair of $\C$ in \eqref{eq:glasso_dual} in each iteration \cite{Banerjee7}.

Our algorithm to solve \eqref{eq:glasso} is thus as follows.
We minimize the GLASSO terms in \eqref{eq:glasso} by solving its dual \eqref{eq:glasso_dual}---iteratively updating one row-column pair of $\C$.
We then project $\C^{-1}$ to $\cH_{\u}^+$ using our projection operator. 
We repeat these two steps till convergence.
Note that in implementation \textit{both steps can be computed using covariance $\C$ directly}, and thus inversion to graph Laplacian $\L = \C^{-1}$ is not necessary until convergence, when we output a solution.

\section{Experimentation}
\label{sec:results}
\subsection{Experimental Setup}
\label{subsec:setup}


We conducted image compression experiments using our modified WebP codec on standard test images from the SIPI Image Database\footnote{https://sipi.usc.edu/database/database.php?volume=misc}.
For the model-based first $K$ eigenvectors in our hybrid transform, we employed the first $K$ frequencies of ADST \cite{han12} (the first $K$ frequencies in a zigzag scan order when computing outer products of 1D ADST frequencies in the horizontal and vertical dimensions, similarly done for 2D DCT).
ADST is well known to perform well for coding of intra-prediction residuals. 
We also followed the practice in \cite{han12} of removing the local mean from the computed prediction residual before transform coding.
The local mean was calculated as the average of the decoded pixels in the $4 \times 4$ sub-block above.

We compared our hybrid transform with fixed transform DCT and adaptive transform KLT, where the latter was composed of eigenvectors computed via the eigen-decomposition of the estimated empirical covariance matrix $\bar{\C}$ as discussed in Section\;\ref{sec:webp}. 
We selected parameter $K$ as $K=1$ and $K=4$ to compare with KLT and DCT on different images. 

We also conducted stability experiments where we varied the number of samples $M$ used to estimate variance $\bar{\C}$, resulting in different estimation reliability. 
The number of $4 \times 4$ sub-blocks used for DC4 mode intra-prediction was roughly $3600$ per image, but varied depending on the image resolution. 

To evaluate the performance of different transforms for different images, we first applied different transform coding schemes to each $4 \times 4$ sub-block chosen for DC4 intra-prediction by WebP. 
For each set of transform coefficients $\balpha$, we normalized its energy to $1$ and sorted them according to energy. 
We plotted the average of cumulative energy percentage for all encoded $4 \times 4$ blocks for each image. 
We also calculated standard deviation (SD) of the cumulative energy at each index and computed the average across the indices to measure the stability of a transform. 

\subsection{Experimental Results} \label{sec:ExResult}

\begin{figure}[t]
\centering
\begin{subfigure}[b]{0.235\textwidth}
\centering
\includegraphics[width=1.85in]{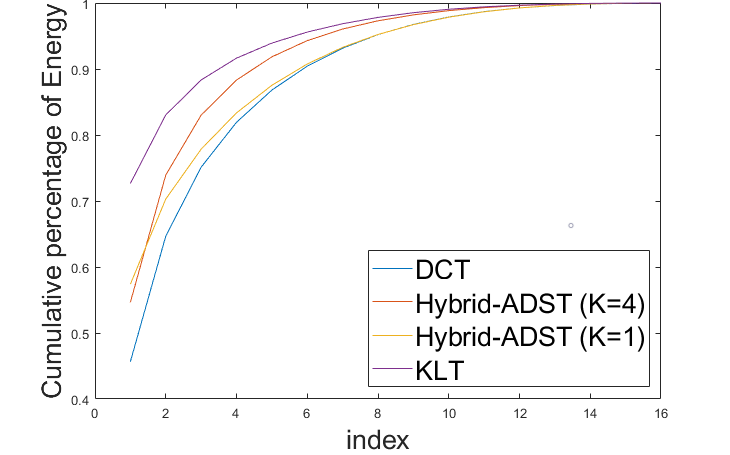}
\caption{\texttt{Airplane}}
\label{fig:Airplane}
\end{subfigure}
\begin{subfigure}[b]{0.235\textwidth}
\centering
\includegraphics[width=1.85in]{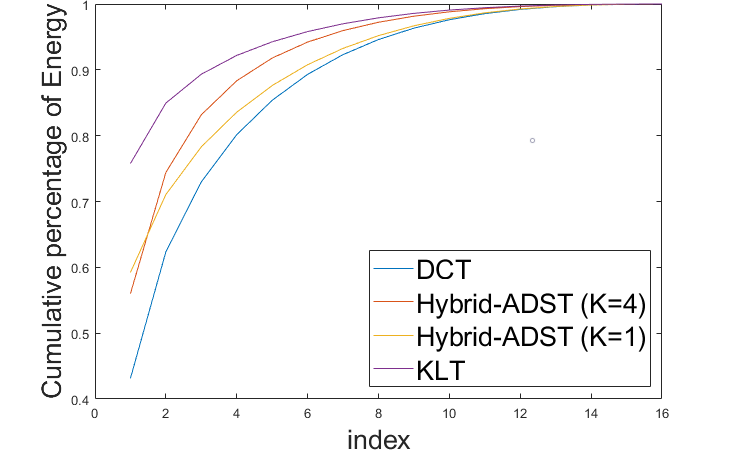}
\caption{\texttt{Pepper}}
\label{fig:Pepper}
\end{subfigure}
\vspace{-0.05in}
\caption{Cumulative energy in percentage vs. number of transform coefficients for test images \texttt{Airplane} and \texttt{Pepper}. Coefficients are sorted in decreasing order of energy. $45$ observations were used to compute empirical covariance $\bar{\C}$.}
\label{fig:EnergyCompaction}
\end{figure}

The plots of cumulative energy versus number of transform coefficients for test images \texttt{Airplane} and \texttt{Pepper} for different transforms are shown in Fig.\;\ref{fig:EnergyCompaction}(a) and (b), respectively.
We observe that both KLT and our proposed Hybrid-ADST performed better than DCT in energy compaction for both \texttt{Airplane} and \texttt{Pepper}. 
This is expected, since adaptive transforms in general perform better than fixed transforms. 
We see also that Hybrid-ADST's performance was between KLT and DCT, since our hybrid transform combines model-based eigenvectors with data-driven eigenvectors.
Moreover, Hybrid-ADST performed better when $K=4$ than when $K=1$.
These results are expected when empirical covariance matrix $\bar{\C}$ is reliable using $M=45$ observations.

\begin{table}[h]
\caption{Average SD of Cumulative Energy (\%) for $M=45$}
\vspace{-0.2in}
\begin{center}
\begin{small}
\begin{tabular}{l|cccr}
Image Name         &\texttt{Couple} &\texttt{Pepper} &\texttt{Airplane}
\\ \hline 
DCT                &5.533  &6.029  &5.652   \\
KLT                &4.483  &5.672  &5.520  \\
Hybrid-ADST ($K=1$)  &5.729  &5.961  &5.676  \\ 
Hybrid-ADST ($K=4$)  &4.475  &5.407  &4.923  \\
\end{tabular}
\end{small}
\end{center}
\label{tab:VarE}
\end{table}

Table\;\ref{tab:VarE} shows the variation of energy compaction for different transforms. 
We see that Hybrid-ADST ($K=4$) had lower average SD than other transforms, meaning that the transform is more stable using the first four ADST frequencies. 
Combining these results with those in Fig.\;\ref{fig:EnergyCompaction}, we can conclude that Hybrid-ADST ($K=4$) offers a relatively good tradeoff between energy compaction and stability compared to other transforms. 

\begin{figure}[t]
\centering
\begin{subfigure}[b]{0.235\textwidth}
\centering
\includegraphics[width=1.85in]{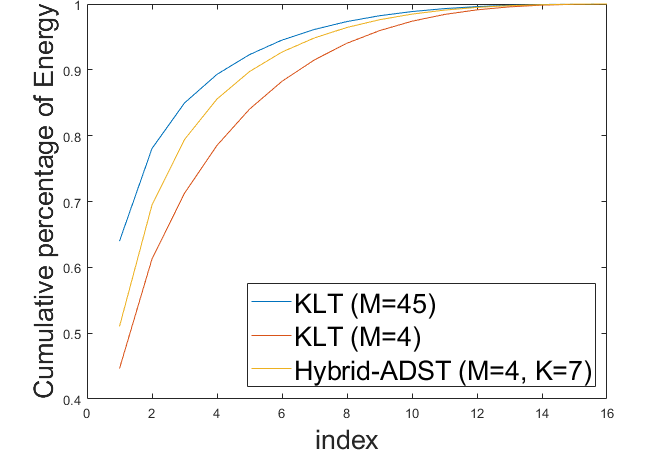}
\caption{\texttt{House}}
\label{fig:House}
\end{subfigure}
\begin{subfigure}[b]{0.235\textwidth}
\centering
\includegraphics[width=1.85in]{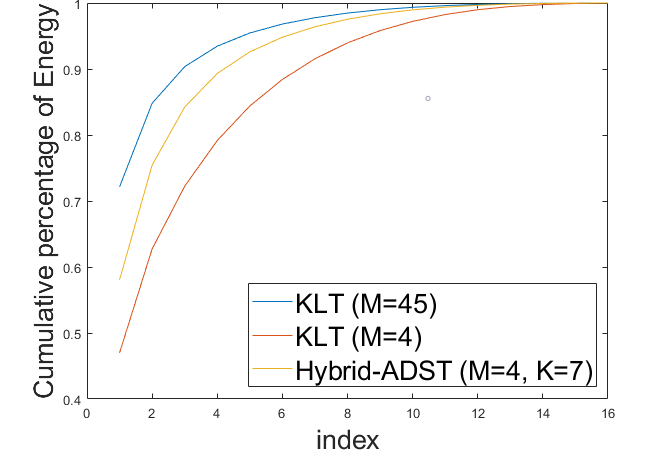}
\caption{\texttt{Couple}}
\label{fig:Couple}
\end{subfigure}
\vspace{-0.05in}
\caption{Cumulative energy in percentage vs. number of transform coefficients for test images \texttt{House} and \texttt{Couple}. Coefficients are sorted in decreasing order of energy. $M=45, 4$ observations were used to compute empirical covariance $\bar{\C}$.}
\label{fig:EnergyCompactionUnreliableC}
\end{figure}

We next varied the number of observations $M$ used to compute variance $\bar{\C}$. 
In Fig.\;\ref{fig:EnergyCompactionUnreliableC}, we observe that KLT's performance dropped significantly when $M$ was reduced from $45$ to $4$, due to the unreliability of estimated $\bar{\C}$. 
In contrast, our proposed Hybrid-ADST ($K=7$) performed well despite the covariance unreliability. 
Table\;\ref{tab:VarEUnreliableC} also shows the stability of Hybrid-ADST ($K=7$) for different images.
This demonstrates the advantage of using model-based eigenvectors to stabilize a transform.

\begin{table}[h]
\caption{Average SD of Cumulative Energy (\%) for different $M$ }
\vspace{-0.2in}
\begin{center}
\begin{small}
\begin{tabular}{l|cccr}
Image Name              &\texttt{Couple} &\texttt{House}  &\texttt{Female}
\\ \hline 
KLT $(M=45)$              &4.843  &6.218  &4.170  \\
KLT $(M=4)$               &7.510  &6.599  &6.225  \\ 
KLT $(M=10)$              &6.277  &6.473  &5.375  \\
Hybrid $(M=4, K=7)$  &4.506  &5.346  &4.639  \\
Hybrid $(M=10, K=7)$ &4.535  &5.260  &4.511  \\
\end{tabular}
\end{small}
\end{center}
\label{tab:VarEUnreliableC}
\end{table}

\section{Conclusion}
\label{sec:conclude}
While fixed transforms like discrete cosine transform (DCT) cannot adapt to non-stationary local image statistics, the adaptive Karhunen-Lo\`{e}ve transform (KLT) is effective only if the empirical covariance matrix $\bar{\C}$ estimated from data is reliable.
In this paper, we proposed a hybrid model-based / data-driven graph transform for image coding, where the first $K$ eigenvectors are derived from a statistical model for stability, while the remaining $N-K$ are computed from $\bar{\C}$ for performance. 
The hybrid graph transform is computed via a graph learning formulation, solved efficiently using augmented Lagrangian relaxation and proximal gradient (PG).
Experimental results show that our hybrid graph transform offers a good tradeoff between energy compaction and error variation.

While we have demonstrated the merits of a hybrid model-based / data-driven transform, the computation of the transform at both the encoder and decoder is expensive.
For future work, we will investigate reduction of computation complexity of such hybrid transform for practical image coding. 

\vfill\pagebreak

\bibliographystyle{IEEEbib}
\bibliography{ref2}

\end{document}